\documentclass[runningheads]{llncs}

\usepackage[T1]{fontenc}
\usepackage{graphicx,verbatim}
\usepackage{amsmath}
\usepackage{xcolor}
\usepackage{url}
\usepackage{hyperref}
\usepackage{xurl}
\usepackage{multirow}
\usepackage{booktabs}
\usepackage{array}
\usepackage[labelfont=bf, labelsep=period]{caption}
\usepackage{makecell}

\begin{document}

\title{Few-shot Deep Learning for Phase-Amplitude Aberration Correction in Transcranial Focused Ultrasound}
\titlerunning{Few-shot Deep Learning for Phase-Amplitude Aberration Correction}

\author{Minju Seol\inst{1} \and Minjee Seo\inst{1} \and Seonaeng Cho\inst{1} \and Kyungho Yoon\inst{1,2}\thanks{Corresponding author.}}
\authorrunning{M. Seol et al.}

\institute{School of Mathematics and Computing (Computational Science and Engineering), Yonsei University, 03722, Seoul, Republic of Korea\\
\email{yoonkh@yonsei.ac.kr}
\and
Innovative \& Intelligent Computational Science Institute (IN2CSI), Seoul, Republic of Korea
}

\maketitle              

\begin{abstract}
Transcranial focused ultrasound (tFUS) is a non-invasive technique that delivers focused acoustic energy through the skull for neuromodulation and therapeutic applications. However, the heterogeneous structure of the skull induces complex, patient-specific phase and amplitude aberrations that distort the acoustic focus and deviate it from the intended target, compromising therapeutic efficacy and safety. Conventional time-reversal (TR) simulations can correct these aberrations but rely on computationally expensive full-wave solvers, making them impractical for real-time use and iterative treatment planning. We propose a few-shot deep surrogate framework that predicts per-element phase and amplitude corrections for a 96-element 3D phased-array transducer from patient CT images. A geometry-aware encoder extracts skull-path features shared across dedicated phase classification and amplitude regression branches, where phase periodicity is handled via circular expectation decoding. The framework is pretrained on diverse skull geometries and fine-tuned with only ten target points, enabling rapid adaptation to unseen patients without full patient-specific simulation. Evaluated via leave-one-out cross-validation across 12 skulls, it achieves a mean phase CMAE of 0.155 rad and amplitude rMAE of 9.089\%, a focal centroid error of 0.467~mm, Dice score of 94.422\%, and peak pressure ratio of 92.332\%, with an approximately 2,535× speedup over TR simulation. The code is available at \href{https://github.com/Minju-Seol/fewshot-tfus-correction}{github.com/Minju-Seol/fewshot-tfus-correction}. 


\end{abstract}

\section{Introduction}
\label{1_Intro}
Transcranial focused ultrasound (tFUS) has emerged as a promising non-invasive technique for thermal ablation, drug delivery, and functional neuromodulation, offering targeted intervention for neurological and psychiatric disorders such as Parkinson's disease, essential tremor, Alzheimer's disease, and depressive disorder~\cite{NEJMoa1300962,Legon2014,Magara2014,tufail2011ultrasonic,yang2025promise}. However, the heterogeneous skull structure distorts the acoustic focus, potentially blurring the focal geometry and deviating it from the intended target, thereby compromising therapeutic effectiveness and safety~\cite{di2019transcranial,jin2020open,zhang2021effects}. To address this, phased-array transducer configurations are widely adopted, enabling electronic manipulation of the phase and amplitude of each element to steer and control the focal spot~\cite{white2005transcranial,yang2025promise}. Yet, because the heterogeneous skull structure induces complex, patient-specific phase delays and attenuation patterns, accurate focusing still requires individualized calibration.

Time-reversal (TR) simulations are conventionally employed~\cite{jing2012time,pernot2007vivo} to correct the skull-induced acoustic distortion, where an acoustic wave emitted from the target point is recorded at each transducer element, then time-reversed and re-emitted to achieve constructive focus at the intended target location~\cite{fink2002time}. This approach enables patient-specific phase and amplitude correction with high accuracy~\cite{wu1992time}. However, as it relies on computationally expensive full-wave numerical solvers such as the pseudo-spectral time-domain (PSTD) method, its applicability in real-time control and iterative treatment planning is limited~\cite{gateau2009transcranial,jin2020open,wang2025systematic}.

Hybrid frameworks integrating physics-based modeling with deep learning have emerged as promising alternatives across domains~\cite{SHIN2024109349,WANG2023107026} to overcome this computational burden. Following this trend, recent studies have applied deep learning to the tFUS domain to accelerate acoustic field simulation and compensate for skull-induced aberrations. However, existing approaches remain limited: Zhang et al.~\cite{zhang2025transcranial} predict skull-induced phase delays but train a separate model for each fixed focal target and condition solely on skull geometry without amplitude correction, limiting scalability across treatment locations and the completeness of aberration compensation. Similarly, Naftchi-Ardebili et al.~\cite{naftchi2026deep} achieve significant computational speedup but operate on 2D acoustic fields without amplitude correction.

To address these limitations, we present a few-shot deep surrogate framework that predicts both the phase and amplitude of each element of a 96-element 3D phased-array transducer in real time. The framework is pretrained on diverse skull geometries and fine-tuned with only 10 target points, enabling rapid adaptation to unseen patients. Our main contributions are: (1) to our knowledge, the first deep learning framework to jointly predict both phase and amplitude for a 3D phased-array transducer, (2) a few-shot adaptation scheme enabling rapid generalization to unseen skulls with only 10 target points, and (3) real-time inference with approximately 2,535$\times$ speedup over TR simulation.

\section{Data Generation}
\label{2_Data}
\subsection{Data Modeling}
\subsubsection{CT-informed Skull Modeling}

Three-dimensional cranial computed tomography (CT) scans were acquired from twelve healthy volunteers using an Aquilion One scanner (Toshiba, Japan). All imaging data were de-identified prior to use, and the study protocol was approved by the institutional review board (IRB). The cranial region relevant to transcranial ultrasound propagation was extracted from the CT scans via intensity-based thresholding and cropping. Each volume was spatially standardized to a uniform field of view and resampled to an isotropic voxel size of 0.5~mm, yielding a $200\times200\times360$ domain used for both model input and ground-truth simulation.

\subsubsection{Phased Array Transducer Configuration}
\label{Phased Array Transducer Configuration}

A 96-element spiral phased array transducer was modeled on a semi-spherical helmet geometry (radius 105~mm, chord length 80~mm), with each element treated as a point monopole source--the standard representation in the k-Wave pseudo-spectral framework~\cite{aubry2022benchmark,treeby2010k}. Element coordinates were defined using the \texttt{makeCartBowl} function in MATLAB k-wave toolbox. The array center was placed at the midpoint of the maximum and minimum y-coordinates of the outermost skull surface, with fine adjustments for consistent placement across subjects. Each element was driven with a 250~kHz continuous wave signal.

\subsubsection{Target Point Sampling}

Target points were defined within a $10\times10\times10~\mathrm{mm}^3$ cube centered at the array center determined in Section~\ref{Phased Array Transducer Configuration}, corresponding to the deep brain region of interest. Points were sampled on a uniform 2~mm grid within this volume, yielding a total of 100 target points. These points serve a dual purpose: as virtual point sources in the TR simulations for extracting ground-truth phase and amplitude values, and as the intended focal targets for phase-amplitude corrected focusing by the phased array transducer.

\subsubsection{Acoustic Path Feature Extraction}

\begin{figure}
    \centering
    \includegraphics[width=0.6\linewidth]{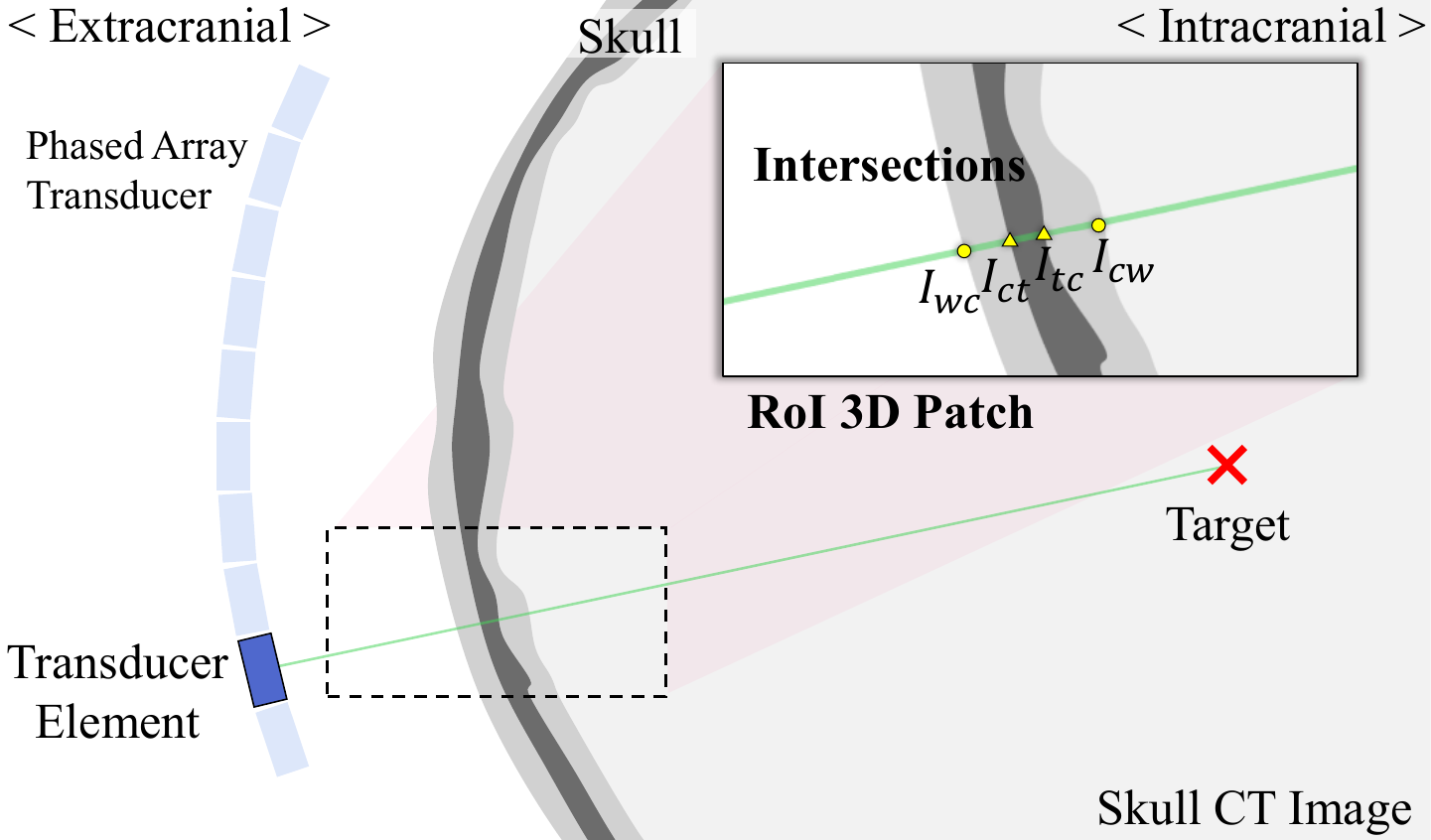}
    \caption{The illustration of Intersection Extraction. The ray from a transducer element to the target defines four skull-interface intersection points that delimit the Region of Interest (RoI) 3D patch.}
    \label{fig_skull_feat}
\end{figure}

For each transducer element, we cast a ray toward the target in CT voxel space and identify four skull-interface intersection points $\{I_{wc}, I_{ct}, I_{tc}, I_{cw}\}$ using adaptive and fixed Hounsfield unit (HU) thresholds, characterizing the layered bone structure traversed by the wave. A fixed-size $40\times40\times80$ region of interest (RoI) patch is then cropped between the outer and inner skull boundaries. This skull RoI patch, the intersection coordinates, and the element and target positions together form the input to the geometry feature encoder (Fig.~\ref{fig_skull_feat}).

\subsection{TR Simulation-based Reference Phase-Amplitude Acquisition}

We adopt TR simulation as our reference method for obtaining ground-truth phase and amplitude values at each transducer element. In TR, a virtual point source placed at the focal target emits a signal that is recorded at each element, time-reversed, and re-emitted, so that the wavefronts retrace their paths and converge at the target, inherently correcting skull-induced distortions without explicit knowledge of the medium~\cite{fink2002time,wu1992time}.

\subsubsection{Simulation Setup and Reference Phase-Amplitude Retrieval}
Acoustic properties were assigned to each voxel based on HU thresholds, categorizing the medium into water/soft tissue ($\mathrm{HU} \le 0$), trabecular bone ($0<\mathrm{HU}<1000$), and cortical bone ($\mathrm{HU} \ge 1000$), with the homogeneous background characterized by $c_0=1500~\mathrm{m/s}$ and $\rho_0=1000~\mathrm{kg/m^3}$; the spatially varying sound speed $c(x)$ and density $\rho(x)$ are given in Table~\ref{tab:acoustic}, which were the values defined from previous studies~\cite{seo2025acoustic,shin2023multivariable,xu2020localized}. These HU-based acoustic properties are consistent with skull-induced phase and amplitude distortions characterized experimentally in our group's prior work using a single-element transducer validated against \textit{ex vivo} calvaria measurements~\cite{jang2025deep,shin2024tfusformer}.

Simulations employed a continuous-wave (CW) excitation at $f_0 = 250~\mathrm{kHz}$, with grid spacing $\Delta x \approx c_0 / (\mathrm{ppw}\cdot f_0)=0.5~\mathrm{mm}$ ($\mathrm{ppw}=12$) and a Courant-Friedrichs-Lewy (CFL) number of 0.3~\cite{5391985}. For each target point, a TR simulation based on the PSTD method was performed using the k-Wave toolbox (v1.4) in MATLAB~\cite{treeby2010k} to obtain the reference phase and amplitude at each transducer element $\textbf{x}_m$ ($m=1,...,96$), via the \texttt{extractAmpPhase} function, yielding the reference pairs $\{A(\textbf{x}_m),\phi(\textbf{x}_m)\}$ that served as training targets.

\begin{table}[t]
\caption{Acoustic properties assigned to each voxel by HU range. Here $x$ denotes the voxel location.}
\centering
\begin{tabular}{|l|r@{\,}c@{\,}l|c|c|}
\hline
Tissue class & \multicolumn{3}{c|}{HU range} & $c(x)$ [m/s] & $\rho(x)$ [kg/m$^3$] \\
\hline
Water/soft tissue   &              & $\mathrm{HU}(x)$ & $\leq 0$      & 1500 & 1000 \\
Trabecular bone     & $0 <$        & $\mathrm{HU}(x)$ & $< 1000$      & 2140 & $1000 + 1.19\,\mathrm{HU}(x)$ \\
Cortical bone       &              & $\mathrm{HU}(x)$ & $\geq 1000$   & 2384 & 2190 \\
\hline
\end{tabular}
\label{tab:acoustic}
\end{table}

\section{Model Architecture}
\label{3_Model}
\begin{figure}
    \centering
    \includegraphics[width=\linewidth]{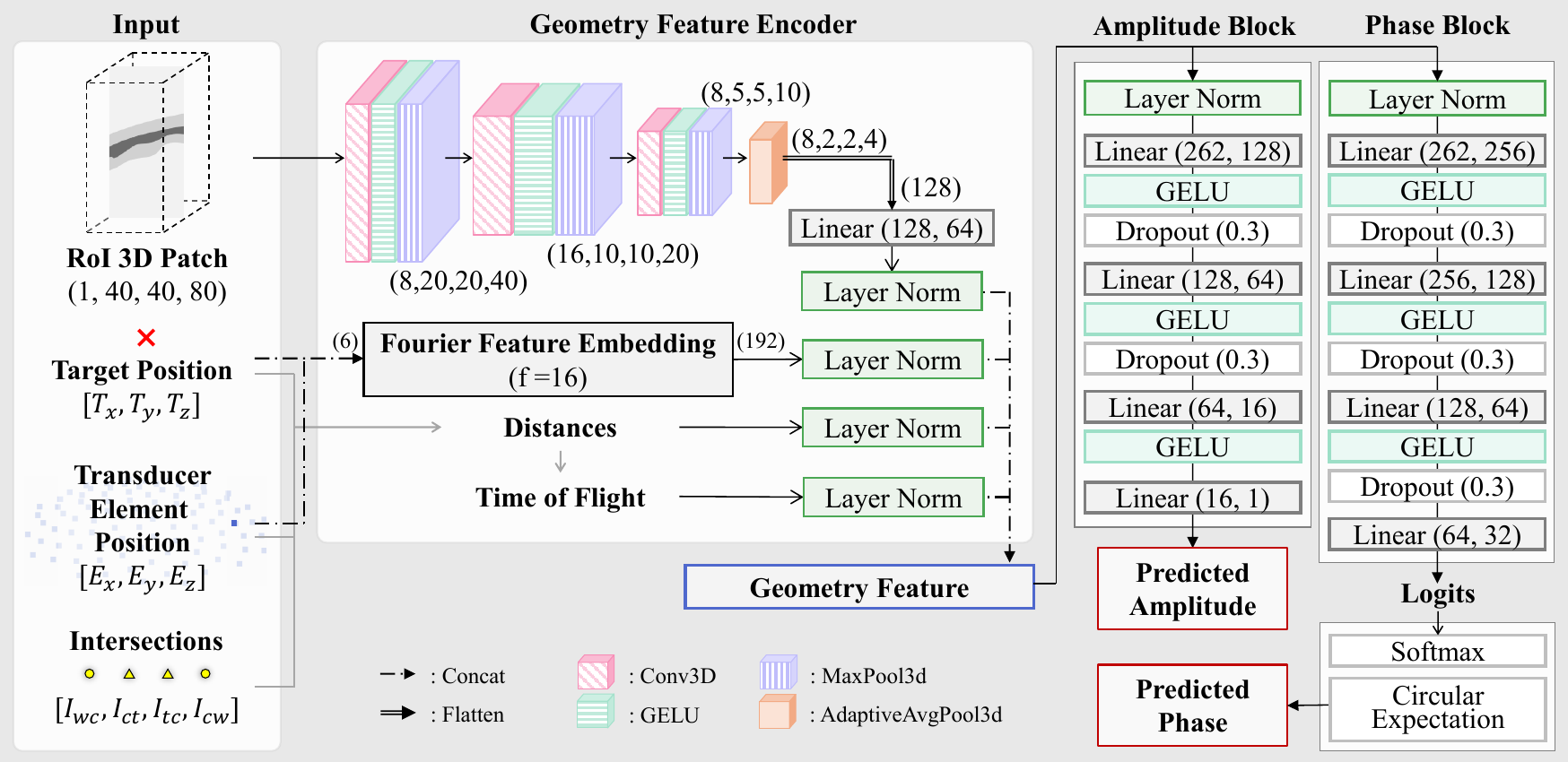}
    \caption{Overview of the proposed network. From the input geometry (RoI patch, target and transducer positions, and skull-path intersections), the geometry feature encoder produces a geometry feature that is fed to the amplitude and phase blocks to predict the per-element amplitude and phase.}
    \label{fig_archi}
\end{figure}

The proposed framework predicts the phase and amplitude that each transducer element should emit to achieve focusing at a desired target location through the skull. The overall pipeline proceeds in two stages: a skull geometry feature embedding stage, followed by element-wise prediction via two downstream models--an amplitude regression model and a phase classification model--each sharing the embedded geometry features.

\subsection{Geometry Feature Encoder}
\label{geom_feat_emb}
The geometry feature encoder transforms the input components--the RoI 3D patch, target and transducer element positions, and path-derived intersections--into a unified geometry feature shared by both downstream models.

The $40\times40\times80$ RoI 3D patch is processed by a 3D convolution network of three Conv3D-GELU-MaxPool3D blocks, adaptively pooled, flattened, and projected to a 64-dimensional vector. In parallel, the target position $[T_x, T_y, T_z]$ and transducer element position $[E_x, E_y, E_z]$ are concatenated and lifted to a 192-dimensional representation via Fourier feature embedding ($f=16$), which mitigates the spectral bias of coordinate-based networks toward low-frequency functions~\cite{Tancik2020-fourier}.

From the four intersection points $\{ I_{wc}, I_{ct}, I_{tc}, I_{cw}\}$, we compute the lengths of five path segments--pre-skull water ($d_{w1}$), cortical, trabecular, and second cortical layers ($d_{c1},d_t,d_{c2}$), and intracranial soft tissue ($d_{w2}$)--and their corresponding time of flight,
$\mathrm{ToF}=(d_{w1}+d_{w2})/c_w+(d_{c1}+d_{c2})/c_c+d_t/c_t,$
with $c_w=1500, c_c=2384,$ and $c_t=2140~\mathrm{m/s}$ for water, cortical, and trabecular bone, respectively. 

The four streams--the convolution patch embedding, Fourier-embedded positions, segment distances, and time of flight--are each normalized by a separate Layer Norm and concatenated into a 262-dimensional geometry feature, shared by the amplitude regression and phase classification models described in Sections~\ref{amp_reg_model} and \ref{ph_class_model}.

\subsection{Amplitude Regression Model}
\label{amp_reg_model}

The amplitude regression model takes the shared 262-dimensional geometry feature and predicts the normalized amplitude assigned to the corresponding transducer element, as shown in Fig.~\ref{fig_archi}. The feature is passed through a layer norm and multilayer perceptron of four linear layers, each intermediate layer using GELU activation and dropout of 0.3, producing a single scalar amplitude. To stabilize regression over the narrow amplitude range, the targets are standardized using the training-set statistics and rescaled by a factor of 5; predictions are mapped back to physical units by the inverse transform at evaluation. The network is trained with the Huber loss.

\subsection{Phase Classification Model}
\label{ph_class_model}

The phase that each transducer element must emit is periodic and wraps discontinuously at the $[-\pi, \pi)$ boundary, which makes direct regression ill-posed near the wrap-around point. We therefore cast phase prediction as a classification problem. The phase range $[-\pi, \pi)$ is discretized into 32 uniform bins, and the model predicts a probability distribution over these classes.

The phase model shares the same 262-dimensional geometry feature and processes it through a layer norm followed by a multilayer perceptron of four linear layers, with GELU activation and dropout of 0.3 between layers, producing 32 class logits. Rather than training against a one-hot target, we use a circular soft label: the ground-truth bin is smoothed over neighboring bins with a Gaussian kernel that respects the periodicity of the phase domain, so that adjacent bins receive graded weight. The network is trained with a Kullback-Leibler divergence loss against this soft target, combined with a circular cosine loss $1-\cos(\phi-\hat{\phi})$ that penalizes angular error directly; the two terms are weighted 0.3 and 0.7, respectively.

At inference, the predicted phase is recovered from the class probabilities by circular expectation, $\hat{\phi}=\mathrm{atan2}(\sum_k p_k \sin \theta_k, \sum_k p_k \cos \theta_k)$, where $p_k$ is the predicted probability of bin $k$ and $\theta_k$ is its center angle. This yields a continuous phase estimate while preserving the circular structure of the output.

\subsection{Model Implementation}

All experiments were conducted on a single NVIDIA GeForce RTX 4090 GPU using PyTorch~\cite{paszke2019pytorch}. Both models are optimized with AdamW (learning rate $5\times10^{-4}$, weight decay $10^{-5}$) and a cosine-annealing schedule (minimum learning rate $10^{-5}$) for 30 epochs with a batch size of 32.

We adopt a leave-one-out (LOO) protocol, where each skull is held out once for testing and the base model is trained on the remaining 11. The base model is then fine-tuned on the held-out skull using only 10 target points with AdamW (learning rate $10^{-4}$, weight decay $10^{-5}$) for 20 epochs, with the 3D convolution encoder frozen to prevent overfitting.

For inference, the geometric features of all 96 elements are fed as a single batch, and both models jointly predict the full steering profile in one forward pass.

\subsection{Evaluation Metrics}
\label{eval_metrics}

We evaluate the proposed framework at two levels: the phase and amplitude prediction accuracy, and the acoustic focus reconstruction performance from those predictions.
We report the circular mean absolute error (CMAE) and the relative mean absolute error (rMAE) for phase and amplitude prediction performance, respectively. 

To assess focusing quality, the predicted phase and amplitude profiles are used to reconstruct the acoustic field, which is compared against the TR ground-truth field. We report three metrics: the focal centroid error (FCE), the peak pressure ratio (PPR), and the focal Dice score. FCE measures the distance between the pressure-weighted centroids of the predicted and TR focal regions, PPR evaluates peak-pressure preservation, and the focal Dice score quantifies the overlap between the predicted and TR focal regions. For FCE and Dice, the focal region is defined as the intracranial full-width-at-half-maximum (FWHM) region, consisting of voxels whose pressure exceeds 50\% of the corresponding peak pressure.

\section{Results}
\label{4_Results}
\subsection{Model Performance}
All results are reported as the mean across 12 LOO trials, with evaluation metrics as defined in Section~\ref{eval_metrics}.

\begin{table}[t]
\caption{Overall performance across 12 skulls. (mean $\pm$ standard deviation (SD))}
\centering
\setlength{\tabcolsep}{7pt}
\begin{tabular}{llc}
\toprule
Class & Metric & mean $\pm$ SD \\
\midrule
\multirow{2}{*}{Prediction Accuracy}
& Phase CMAE [rad] $\downarrow$ & 0.155 $\pm$ 0.082 \\
& Amplitude rMAE [\%] $\downarrow$ & 9.089 $\pm$ 0.720 \\
\midrule
\multirow{3}{*}{Focusing Performance}
& FCE [mm] $\downarrow$ & 0.467 $\pm$ 0.317 \\
& Dice [\%] $\uparrow$ & 94.422 $\pm$ 2.796 \\
& PPR [\%] $\uparrow$ & 92.332 $\pm$ 6.132 \\
\bottomrule
\end{tabular}
\label{tab:results}
\end{table}

\subsubsection{Prediction Accuracy}

The model achieved a mean Phase CMAE of 0.155 rad and a mean Amplitude rMAE of 9.089\%, as summarized in Table~\ref{tab:results}. Across individual skulls, Phase CMAE ranged from 0.081 rad (Skull 5) to 0.336 rad (Skull 4), reflecting moderate variability attributable to differences in skull geometry and thickness. Amplitude rMAE remained relatively stable across skulls, ranging from 8.141\% (Skull 5) to 11.015\% (Skull 3), suggesting that amplitude prediction was consistently achieved regardless of inter-subject anatomical variation.

\subsubsection{Focusing Performance}
\begin{figure}[t]
    \centering
    \includegraphics[width=0.6\linewidth]{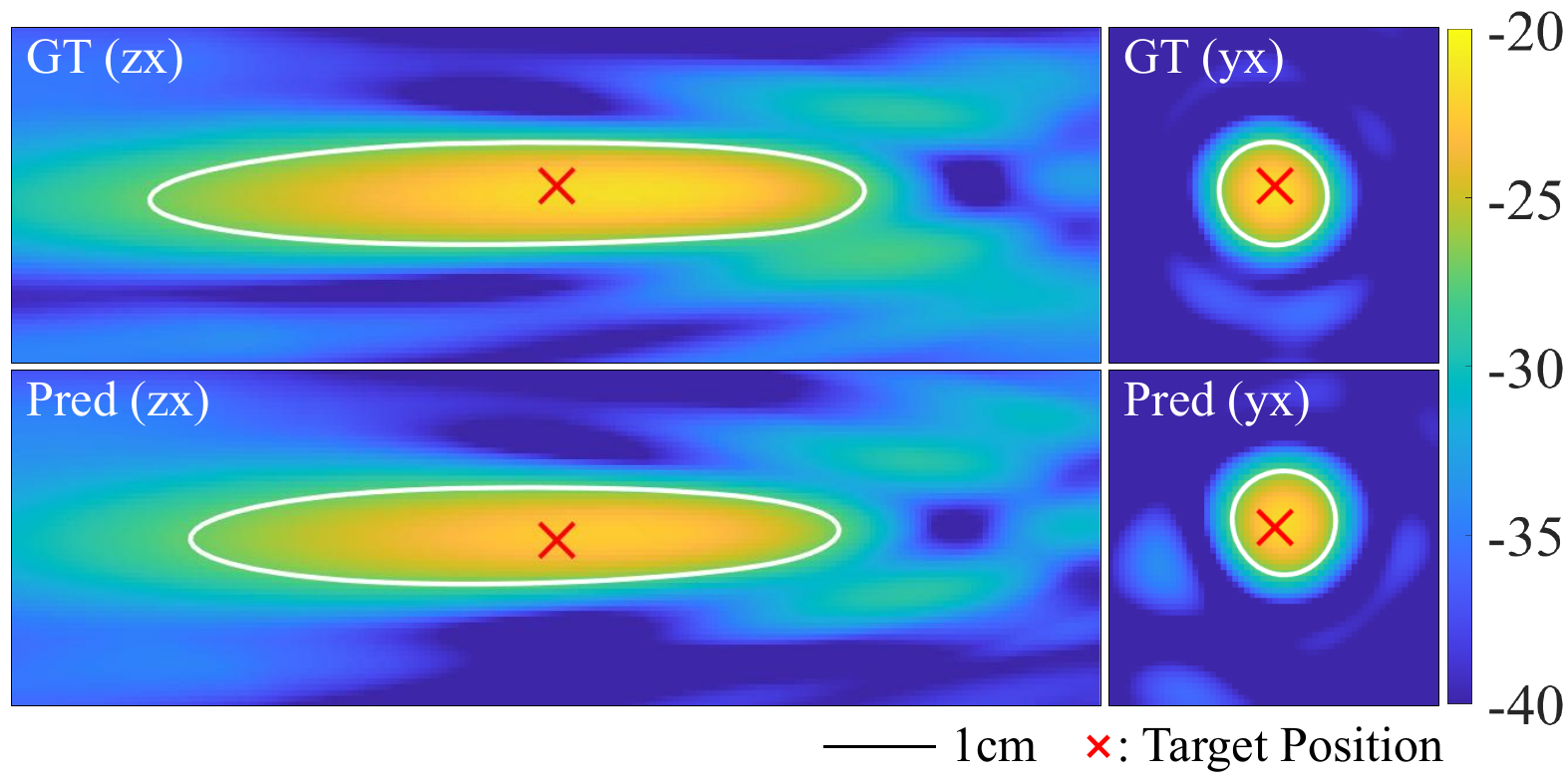}
    \caption{Pressure fields for a representative case (Skull 9, Target Point 66) in the zx- and yx-planes. GT (top) and Pred (bottom) with $-6$ dB FWHM contour (white) and target position (red cross).}
    \label{fig:pmap}
\end{figure}

The proposed framework achieved a mean FCE of 0.467~mm, Dice score of 94.422\%, and PPR of 92.332\%, as summarized in Table~\ref{tab:results}. Per-skull FCE ranged from 0.084 mm (Skull 5) to 1.200 mm (Skull 4), while Dice remained above 89\% across all skulls. To qualitatively visualize this performance, Fig.~\ref{fig:pmap} presents the reconstructed pressure fields for Skull 9, Target Point 66, selected as the most representative among all 12$\times$90 test cases by minimum normalized Euclidean distance to the mean across all five metrics.

\subsubsection{Real-time Applicability}
The proposed framework computes the complete phase and amplitude profile across all 96 elements in 0.029~s per target, compared with 73.52~s for TR simulation--an approximately 2,535$\times$ speedup--while few-shot adaptation to an unseen skull completes in 17.29~s.

\subsection{Component Contribution Analysis}

Table~\ref{tab:ablation} summarizes the component contribution test results for two key design choices, evaluated on phase prediction accuracy. Removing Fourier feature embedding and replacing it with raw coordinates led to a dramatic increase in Phase CMAE from 0.155 to 0.955 rad, demonstrating that high-frequency positional encoding is essential for capturing spatial dependencies between transducer elements and target positions. Replacing the phase classification head with direct regression also degraded performance (0.282 rad), validating the use of circular expectation decoding for handling phase periodicity. These results confirm that both Fourier feature embedding and phase classification with circular expectation decoding are essential components of the proposed framework.

\begin{table}[t]
\caption{Component contribution test on key design choices evaluated on phase prediction accuracy (mean $\pm$ SD across 12 skulls).}
\centering
\setlength{\tabcolsep}{7pt}
\begin{tabular}{lc}
\toprule
Variant & CMAE [rad] $\downarrow$ \\
\midrule
w/o Fourier embedding & 0.955 $\pm$ 0.099 \\
w/ phase regression   & 0.282 $\pm$ 0.037 \\
\textbf{Ours}         & \textbf{0.155 $\pm$ 0.082} \\
\bottomrule
\end{tabular}
\label{tab:ablation}
\end{table}

\section{Conclusion}
\label{5_Conclusion}
We presented a few-shot deep surrogate framework for real-time transcranial phase and amplitude aberration correction in tFUS, achieving an approximately $2{,}535\times$ speed-up over TR simulation with rapid adaptation to unseen skulls using only ten target points. The physics-informed geometry encoder with Fourier-embedded positional features enables accurate and generalizable prediction of both phase and amplitude steering parameters across diverse skull geometries. Future work will explore diverse transducer configurations and variable target trajectories, incorporate larger and more heterogeneous patient cohorts, and ultimately move toward experimental phantom studies and clinical validation. 

\bibliographystyle{splncs04}
\bibliography{_References}

\end{document}